\documentclass[preprint2]{aastex}

\usepackage{amsmath}
\usepackage{latexsym}
\usepackage{amssymb}
\usepackage{url}
\usepackage{graphicx}
\usepackage{ulem}
\usepackage{color,soul}

\newcommand{\grad}{\mathbf{\nabla}}

\shorttitle{Solar $p$-Mode Driven Sausage Waves}
\shortauthors{Gascoyne et al.}

\begin{document}


\title{Energy Loss of Solar $p$ Modes due to the excitation of Magnetic Sausage Tube Waves: Importance of Coupling the Upper Atmosphere}


\author{A. Gascoyne$^1$, R. Jain$^1$ and B. W. Hindman$^2$}
\affil{$^1$Applied Mathematics Department, University of Sheffield,
    Sheffield, S3 7RH, UK}
\email{a.d.gascoyne@sheffield.ac.uk, r.jain@sheffield.ac.uk}
\affil{$^2$JILA, University of Colorado at Boulder, USA}

\begin{abstract}
We consider damping and absorption of solar $p$ modes due to their energy loss to magnetic tube waves that can freely carry energy out of the acoustic cavity.  The coupling of $p$ modes and sausage tube waves is studied in a model atmosphere composed of a polytropic interior above which lies an isothermal upper atmosphere.  The sausage tube waves, excited by $p$ modes, propagate along a magnetic fibril which is assumed to be a vertically aligned, stratified, thin magnetic flux-tube.  The deficit of $p$-mode energy is quantified through the damping rate, $\Gamma$ and absorption coefficient, $\alpha$.  The variation of $\Gamma$ and $\alpha$ as a function of frequency and the tube's plasma properties is studied in detail.\\
\\
Previous similar studies have considered only a subphotospheric layer, modelled as a polytrope that has been truncated at the photosphere \citep{1996BogdanHCC,2008HindmanJ,2011GascoyneJH}. Such studies have found that the resulting energy loss by the $p$ modes is very sensitive to the upper boundary condition, which because of the lack of a upper atmosphere have been imposed in a somewhat ad hoc manner. The model presented here avoids such problems by using an isothermal layer to model the overlying atmosphere (chromosphere), and consequently, allows us to analyse the propagation of $p$-mode driven sausage waves above the photosphere.  In this paper we restrict our attention to frequencies below the acoustic cut-off frequency.  We demonstrate the importance of coupling all waves (acoustic, magnetic) in the subsurface solar atmosphere with the overlying atmosphere in order to accurately model the interaction of solar $f$ and $p$ modes with sausage tube waves.  In calculating the absorption and damping of $p$ modes we find that for low frequencies, below $\approx3.5$ mHz, the isothermal atmosphere, for the two region model, behaves like a stress-free boundary condition applied at the interface ($z=-z_{0}$).
\end{abstract}



\section{Introduction}
The turbulent motions in the upper convection zone of the Sun are thought to be the source of acoustic energy, stochastically exciting $p$ modes.  The convection zone is also the region where many thin magnetic fibril fields reside, continually shaken by these modes.\\
\\
Solar $p$ modes have been studied extensively using high-quality data gathered by ground and spaced based telescopes, namely at the six BiSON sites \citep{1997Chaplin}, the GONG network \citep{1996HillSSA}, the GOLF experiment and MDI instrument aboard SOHO \citep{1997LazrekBBB,1995ScherrerBB}, and more recently HMI aboard SDO \citep{2012SchouSB}.  We now have a clear and consistent picture of the observed linewidths (damping rates) associated with $f$ and $p$ modes \citep{1997ChaplinEIMMN,1997LazrekBBB,2000KommHHa,2000KommHHb,2001KommHH}.  The most notable feature of the observed damping rates is the `plateau' between $2.5$ and $3.1$ mHz followed by a gradual rise.  Damping rates indicate a global observational signature of the energy loss experienced by the $p$ modes while a local signature of this energy loss is characterised by absorption coefficient which is a measure of the difference between the incoming and outgoing $p$-mode power around a magnetic region.  Magnetic features observed within the solar photosphere such as sunspots, plages and pores are efficient absorbers of $p$-mode energy.  The absorption coefficient was first measured by a Fourier-Hankel decomposition method \citep{1988BraunDL,1995Braun} and in recent years by helioseismic holography \citep{2008BraunB}.  The frequency dependence of the absorption coefficient has a distinctive bell curve structure which is consistent over a variety of magnetic regions and $p$-mode orders; with a peak between $3$$-$$4$ mHz.  The saturation of the absorption coefficient seen above $4$ mHz is likely a result of the acoustic emission of $p$-mode power observed around magnetic features at higher frequencies \citep[see e.g.,][]{2002JainH}.\\
\\
There have been several theoretical studies to explain the observed frequency dependence of damping rates and absorption coefficients.  \citet{1996BogdanHCC} developed a model whereby $p$-mode energy is lost to magnetic flux tubes via the excitation of MHD tube waves by $p$-mode buffeting.  They found that tube mode excitation on fibril magnetic fields may be a dominant source of low-frequency $p$-mode damping rates.  Further work on this mechanism of $p$-mode damping by \citet{2008HindmanJ} found that the damping rates are very sensitive to the upper boundary condition.  They also demonstrated that $p$ modes efficiently generate tube waves and that an energy flux in excess of $10^5$ ergs cm$^{-2}$ s$^{-1}$ can be driven upward through photospheric levels.  \citet{2009JainHBB} presented a method for calculating the absorption coefficient for this mechanism for a single magnetic flux tube and for a collection of identical tubes to simulate a plage region observed by \citet{2008BraunB}.  This work was extended to incorporate a fixed and random distribution of tubes with varying plasma properties, thus enabling us to measure and estimate the statistical properties of flux tubes within an observed plage region \citep[see][for details]{2011GascoyneJH,2011JainGHa,2011JainGHb}.\\
\\
All the above mentioned studies modelled the magnetic fibrils and their surroundings with a truncated polytrope.  Regardless of whether a stress-free \citep{1996BogdanHCC} or a maximal-flux \citep{2008HindmanJ} boundary condition was assumed at the truncation height \citep[see also][for a radiation boundary condition]{1999CrouchC}, the models did not represent wave reflection from the upper photosphere/lower chromosphere satisfactorily.  Also, these studies revealed that the magnitude of damping rate, $\Gamma$, and absorption coefficient, $\alpha$, were sensitive to the boundary conditions used.  Here, we calculate both damping rates and absorption coefficients for the equilibrium model where the magnetic fibril is extended into an isothermal region above a polytrope.  We demonstrate that the coupling of sausage waves and external $p$ modes to the overlying atmosphere is necessary to avoid the sensitivity of boundary conditions applied at the model photosphere ($z=-z_{0}$).\\
\\
The paper is organised as follows.  In \S \ref{equi:conf} we describe the equilibrium atmosphere of our two-region model and the static configuration of the magnetic flux tubes.  Section \ref{wavefield:nm} details the wavefield in the non-magnetic surroundings for both the polytropic and isothermal atmospheres separately, and the coupling conditions across the interface.  The wavefields describe the propagating (in the polytrope) and evanescent (in the isothermal region) form of the $f$ and $p$ mode oscillations.  In \S \ref{sausage:excite} we derive the driven sausage waves of the thin magnetic flux tube and calculate the energy flux of sausage waves at the bottom and top of the polytropic and isothermal atmosphere respectively.   In \S \ref{results:discuss} we present the damping rates, absorption coefficients of $p$ modes and discuss our results.
\begin{figure}[t]
\epsscale{.80}
\plotone{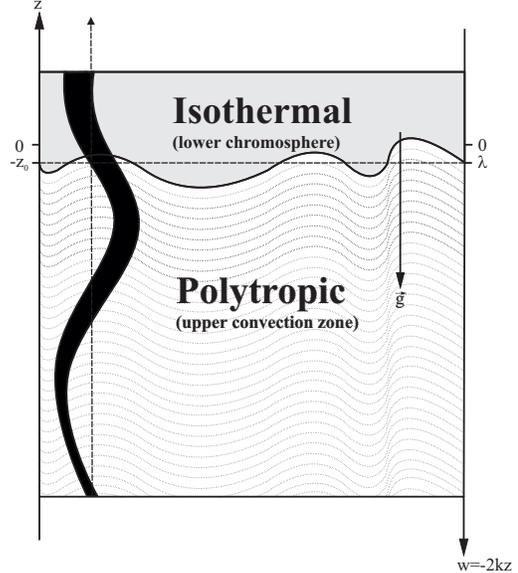}
\caption{Cartoon of the proposed model configuration where the right axis is the non-dimensional representation of the left axis and consequentially points in the opposite direction.  The thin magnetic flux tube undergoes wave-like behaviour due to the motions of the surrounding medium.  The interface resides at $z=-z_{0}$, where we apply the continuity of the Lagrangian pressure perturbation; this explains the non-ridged nature of this boundary and why the two regions interact non-linearly.}\label{fig:atmos}
\end{figure}

\section{Equilibrium configuration}\label{equi:conf}
In this section we describe the equilibrium model of a slender magnetic fibril as an axisymmetric magnetic flux tube embedded in a plane-parallel, gravitationally stratified two-region atmosphere.  The lower and upper layers of the model represent the subsurface and the upper atmosphere respectively.   It is therefore appropriate to consider polytropic and isothermal atmospheres for the interior and the upper photosphere respectively.  A schematic representation of the excitation of tube waves along a thin magnetic fibril due to $p$-mode buffeting in our two-region model is shown in figure \ref{fig:atmos}.\\
\\
The surrounding nonmagnetic medium is modeled as a polytropic atmosphere below the isothermal atmosphere.  The vertical variation of the pressure $p_{e}$ and density $\rho_{e}$ for the polytropic atmosphere is,
\begin{equation}\label{poly:equi}
p_{e}=p_{0}\left(-\frac{z}{z_{0}}\right)^{a+1},\quad\rho_{e}=\rho_{0}\left(-\frac{z}{z_{0}}\right)^{a},
\end{equation}
where $\rho_{0}$ is the density at the interface $z=-z_{0}$ and $p_{0}=gz_{0}\rho_{0}/(a+1)$.  The isothermal atmosphere which represents the upper photosphere/lower chromosphere of the Sun has it's lower boundary at the truncation depth $z=-z_{0}$ and extends infinitely upward as $z\rightarrow\infty$.  The vertical variation of the pressure $p_{e}$ and density $\rho_{e}$ in the isothermal atmosphere is,
\begin{equation}\label{iso:equi}
p_{e}=p_{0}e^{-(z+z_{0})/H},\quad\rho_{e}=\rho_{0}e^{-(z+z_{0})/H},
\end{equation}
where the scale height $H=z_{0}/(a+1)$.  The sound speed for both atmospheres is $c^2=\gamma p_{e}/\rho_{e}$.\\
\\
It is clear from equations (\ref{poly:equi}) and (\ref{iso:equi}) that the equilibrium quantities are continuous through the interface.  In the polytropic atmosphere the pressure, density and sound speed vary as a power law where $a$ is the polytropic index and is related to the ratio of specific heats $\gamma$ via $\gamma=(a+1)/a$.  In the isothermal atmosphere the pressure and density decrease exponentially with height at the same rate, thus the sound speed (and the temperature) is constant.\\
\\
The characteristic physical scales for our model photosphere ($z=-z_{0}$) are assumed to coincide with the photospheric reference model of \citet{1986MaltbyACKKL}, and so we adopt $\rho_{0}=2.78\times 10^{-7}$ g cm$^{-3}$, $p_{0}=1.21\times 10^5$ g cm$^{-1}$ s$^{-2}$, and $g=2.775\times10^4$ cm s$^{-2}$.  The choice of polytropic index $a=1.5$ yields the truncation depth $z_{0}=392$ km and a photospheric sound speed of $8.52$ km s$^{-1}$.\\
\\
Now we define the profile of our magnetic fibril field, first introduced by \citet{1996BogdanHCC} \citep[see also][]{1997Hasan,2008HindmanJ,2009JainHBB,2011GascoyneJH}.  We assume the magnetic fibril is a thin, untwisted, axisymmetric and vertically aligned magnetic flux tube which lacks internal lateral structure.  For such a thin magnetic flux tube to be in thermal equilibrium and hydrostatic balance requires that temperature and the total pressure is uniform across the tube and is equal to the external pressure and temperature.  Thus the magnetic pressure has the same scale height as the gas pressure which results in a constant plasma-$\beta$ (ratio of gas to magnetic pressure) with height within the tube.  Since the gas pressure decreases rapidly with height in the solar atmosphere so does the magnetic pressure; this then results in significant flaring of field lines which inevitably will break the thin flux tube approximation.  However, if the skin depth of the wave solutions are sufficiently small, the waves will not sense the uppermost portion of the atmosphere where the approximation becomes invalid.  Therefore, we can safely ignore this problem if we restrict our attention to frequencies below the cut-off frequency for the isothermal atmosphere.  The lateral variation of the magnetic field is ignored \citep[see for details]{1996BogdanHCC}.  Thus the tube's internal pressure $P(z)$, density $\rho(z)$, magnetic field $B(z)$, and cross-sectional area $A(z)$ can be described as follows,
\begin{equation}\label{back:p}
    p(z)=\frac{\beta}{\beta+1}p_{e}(z),
\end{equation}
\begin{equation}\label{back:den}
    \rho(z)=\frac{\beta}{\beta+1}\rho_{e}(z),
\end{equation}
\begin{equation}\label{B}
    \frac{B^2(z)}{8\pi}=\frac{1}{\beta+1}p_{e}(z),
\end{equation}
\begin{equation}\label{CArea}
    A(z)=\frac{\Theta}{B(z)}=\left(\frac{\beta+1}{8\pi p_{e}(z)}\right)^{1/2}\Theta,
\end{equation}
where $\Theta$ is the constant magnetic flux.  All parameters associated with the external non-magnetic atmosphere which have an internal counterpart associated with the magnetic flux tube will be distinguished with the subscript `$e$'.
\begin{figure}[h]
\epsscale{.80}
\plotone{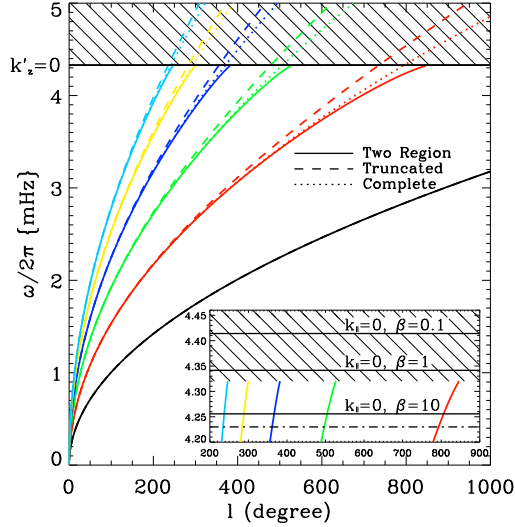}
\caption{Dispersion relation for non-magnetic waves in the two-region model (solid) considered here, and previously studied truncated (dashed) and complete (dotted) polytropes.  Each mode order is plotted with a different colour: $f$ (black), $p_{1}$ (red), etc.  In this paper we restrict our attention to frequencies below the cut-off (indicated by the horizontal solid line).  Inset plot shows the location of the sausage wave cut-off frequency ($k_{\|}=0$, solid horizontal line) for $\beta=0.1$, $1$, $10$.  Shaded region denotes the zones of propagation for $p$ modes in the isothermal atmosphere (of which we do not consider).  The dot-dashed line indicates the frequency limit where the upper boundary condition is valid i.e., $|k_{z}'|+|k_{\|}|>\frac{1}{4H}$.}\label{fig:disp}
\end{figure}

\section{Wavefields in the field-free atmosphere}\label{wavefield:nm}
\begin{figure}[t]
\epsscale{1.0}
\plotone{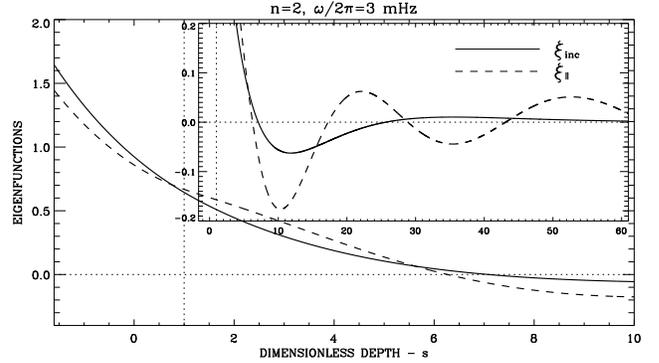}
\caption{Vertical displacements for the incident $p_{2}$-mode, $\xi_{\text{inc}}$, (solid line) and corresponding sausage wave solution, $\xi_{\|}$, (dashed line) for $\beta=1$.  The vertical dotted line denotes the interface at $s=1$.  Inset plot shows the propagating portion of the wave solutions in the polytropic region.}\label{fig:eigenf3n2b1}
\end{figure}

We assume that the fluid displacement takes the form,
\begin{equation}
\vec{\xi}_{e}=\tilde{\xi}_{e}(z)e^{i(kx-\omega t)},
\end{equation}
where $\tilde{\xi}_{e}(z)$ is the vertical eigenfunction of the displacement vector.  For the polytrope, the temperature varies linearly with depth ($\propto -gz/a$) thus the vertical eigenfunction is represented by the Whittaker function $W_{\kappa,\hspace{0.5mm}\mu}$ (see equation (\ref{wave:sol:poly})).  Whereas, for the isothermal region the temperature is constant and so we also Fourier analyse the governing equation (equation (\ref{iso:gov})) in height.  Thus, the vertical wavenumber in the isothermal region is defined as,
\begin{equation}\label{kz'}
k'_{z}=\sqrt{\left(\omega^2-\omega^2_{\text{ac}}\right)\frac{1}{c^2}-\left(\omega^2-\omega^2_{\text{\tiny BV}}\right)\frac{k_{x}^2}{\omega^2}}.
\end{equation}
where $\omega_{\text{ac}}=\gamma g/(2c)$ and $\omega_{\text{\tiny BV}}=\sqrt{\gamma-1}g/c$ are the acoustic cut-off and Brunt V\"{a}is\"{a}l\"{a} frequencies respectively.  We match the wave perturbations in the two atmospheres by ensuring continuity of the vertical displacement $\xi_{z,e}$ and Lagrangian pressure perturbation $\delta p_{e}$ across the interface $z=-z_{0}$.  The upper and lower boundary conditions are that the wave energy density vanishes as $z\rightarrow\pm\infty$.  The details of this procedure is described in Appendix \ref{A:wavefield:nm}.\\
\\
In order to satisfy the matching conditions we solve equation (\ref{freq:eigen}) for $\kappa$, at each dimensionless frequency $\nu$.  We obtain a discrete set of eigenvalues which correspond to each individual wave mode (figure \ref{fig:disp}) similarly to the truncated \citep{1996BogdanHCC} and complete \citep{1995FanBC} polytrope cases.  The fundamental mode ($n=0$) is the same for all three cases which is expected since this mode is a surface gravity wave i.e., incompressible ($\grad\cdot\vec{\xi_{e}}=0$) in both the polytropic and isothermal atmospheres.  The isothermal atmosphere sets a global frequency cut-off when $k'_{z}=0$ (see equation (\ref{kz'})).  For a polytropic index of $a=1.5$ and an interface located at $z=-z_{0}$ where $z_{0}=392$ km, the cut-off frequency is $\approx 4.3$ mHz (figure \ref{fig:disp}), below this frequency the waves are evanescent whereas above the cut-off the waves are propagating in the isothermal region.  In this paper we are concerned with trapped $p$-modes thus we calculate eigenvalues and eigenfunctions for the two-region model for frequencies $\omega/(2\pi)\leq 4.3$ mHz.  The solid curve in figure \ref{fig:eigenf3n2b1} illustrates the vertical displacement for the $p_{2}$-mode at $3$ mHz, clearly showing the propagating portion of the wave in the polytropic region and the evanescent tail in the isothermal region.

\section{Excitation of sausage waves}\label{sausage:excite}
The incident $f$ and $p$ modes, described by equations (\ref{xi:iso}) and (\ref{xi:poly}), drive waves along the thin magnetic tube, thus removing energy from the $p$-mode cavity.  Here we shall concentrate solely on the excitation of sausage waves which can be described as an axisymmetric pressure driven wave with motions parallel to the magnetic flux tube.  Under the thin flux tube approximation we derive a set of MHD equations for an inviscid and infinitely conducting medium.  On linearising these equations and assuming a time dependence consistent with the external perturbations we can describe the interaction of the field-free medium with the thin flux tube for the sausage mode in both the polytropic and isothermal atmosphere.  Thus using the formulation of \citet{2008HindmanJ,2009JainHBB,2011GascoyneJH} for the polytropic atmosphere and \citet{1997Hasan} for the isothermal atmosphere we obtain the governing equations for the propagation of sausage waves within our prescribed two-region atmosphere,
\begin{figure*}[t]
\centering
\includegraphics[width=15cm]{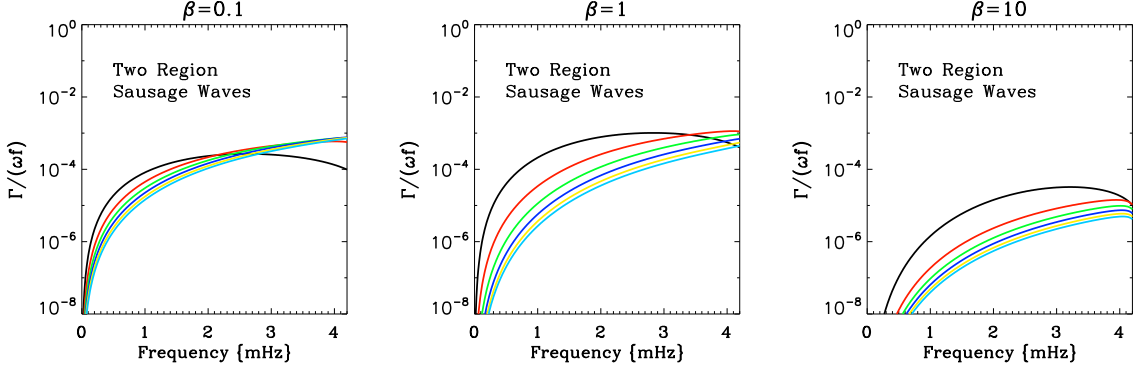}
\caption{Damping rate of $f$ and $p$ modes due to excitation of sausage waves on thin magnetic flux tubes embedded in a two-region polytropic-isothermal atmosphere.  Each mode order is plotted with a different colour, see figure \ref{fig:disp} for the colour scheme.}\label{fig:damp4H}
\end{figure*}
\begin{figure*}[t]
\centering
\includegraphics[width=15cm]{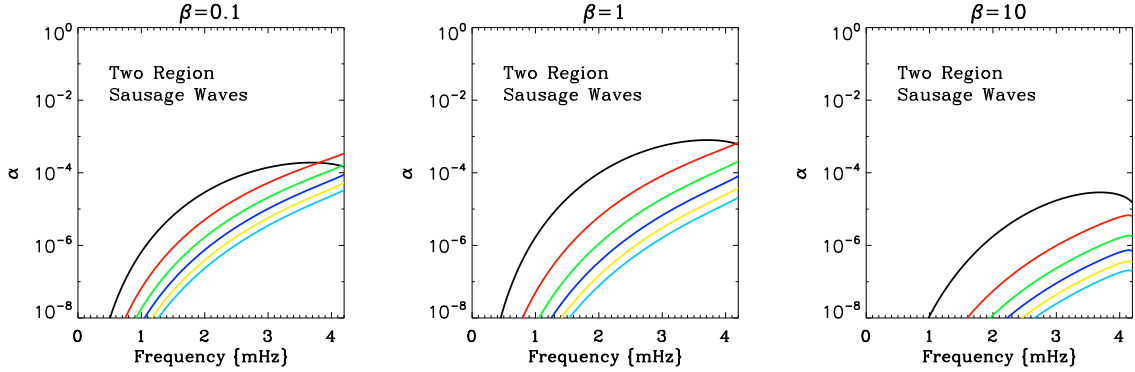}
\caption{Absorption coefficient for a single tube embedded in a two-region polytropic-isothermal atmosphere.  Each mode order is plotted with a different colour, see figure \ref{fig:disp} for the colour scheme.}\label{fig:alpha4H}
\end{figure*}
\begin{figure}[t]
\epsscale{.80}
\plotone{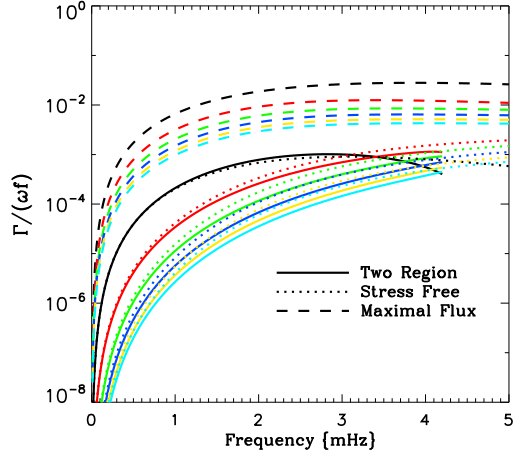}
\caption{Damping rate of $f$ and $p$ modes due to the excitation of sausage waves on thin magnetic flux tubes embedded in a two-region polytropic-isothermal atmosphere for $\beta=1$.  Height at which we calculate $E^{(u)}_{\|}$ is, as $z\rightarrow\infty$ scale heights above the interface.  Overplotted are the cases of applying a stress-free (dotted) and maximal-flux (dashed) boundary condition at the interface ($z=-z_{0}$) \citep[][]{1996BogdanHCC,2008HindmanJ}.  Each mode order is plotted with a different colour, see figure \ref{fig:disp} for the colour scheme.}\label{fig:dampCz0392}
\end{figure}
\begin{equation*}
\left[c^2_{\text{\tiny T}}\frac{\partial^2}{\partial z^2}-\frac{\gamma g}{2}\frac{c^2_{\text{\tiny T}}}{c^2}\frac{\partial}{\partial z}+\omega^2+N_{g}\right]\xi_{z}
\end{equation*}
\begin{equation}\label{gov:eqn}
\hspace{4cm}=\frac{c^2_{\text{\tiny T}}}{\rho V^2_{\text{\tiny A}}}\left(\frac{\partial}{\partial z}+\frac{g}{c^2}\right)p_{e}',
\end{equation}
where $V^2_{\text{\tiny A}}=B^2/(4\pi\rho)$ and $c^2_{\text{\tiny T}}=c^2V^2_{\text{\tiny A}}/(1+c^2/V^2_{\text{\tiny A}})$ are the Alfv\'{e}n speed and tube speed respectively.  The governing equations for the propagation of sausage waves in the two regions are essentially the same apart from the term $N_{g}$ ($=-\omega_{\text{\tiny BV}}^2\left(c^2+\frac{\gamma}{2}V^2_{\text{\tiny A}}\right)/(c^2+V^2_{\text{\tiny A}})$) which permits the propagation of gravity waves in the isothermal layer; this quantity is zero for the polytrope which is in adiabatic equilibrium.\\
\\
The solutions to equation (\ref{gov:eqn}) are derived in detail in Appendix \ref{GE:sausage} and take the form,
\begin{equation}
\vec{\xi}_{\|}=\xi_{\|}(z)e^{-i\omega t}\hat{z},
\end{equation}
where $\xi_{\|}(z)$ is written explicitly in equations (\ref{xi:I}) and (\ref{xi:P}) for the isothermal and polytropic atmospheres respectively.  The solution in the polytropic atmosphere is a combination of Hankel functions and Greens functions, whereas in the isothermal region the solution can be written in terms of Fourier components, thus the vertical wavenumber for the sausage wave in the isothermal atmosphere can be written as,
\begin{equation}
k_{\|}=\sqrt{\frac{\omega^2-\omega_{s}^2}{c^2_{\text{\tiny T}}}},\quad\omega_{s}^2=\omega_{\text{\tiny BV}}^2+\frac{c^2_{\text{\tiny T}}}{H^2}\left(\frac{3}{4}-\frac{1}{\gamma}\right)^2.
\end{equation}
The matching conditions for the sausage tube wave, applied at $z=-z_{0}$, are analogous to the field-free medium in that the continuity of the vertical displacement $\xi_{\|}$ and Lagrangian pressure perturbation $\delta p_{\|}$ across the interface between the two regions are ensured (discussed in detail in \S\ref{match:sausage}).  The upper and lower boundary conditions we require are that the wave energy density is finite as $z\rightarrow\pm\infty$.  The details are given in Appendix \ref{GE:sausage}.

\section{Results and discussion}\label{results:discuss}
\begin{figure}[t]
\epsscale{.80}
\plotone{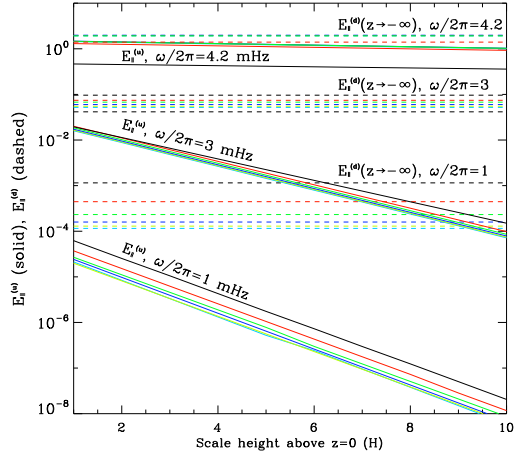}
\caption{Upward sausage wave energy flux, $E_{\|}^{(u)}$, as a function of height (solid) for $\beta=1$ and $\omega/2\pi=1$, $3$ and $4.2$ mHz (labelled accordingly).  Also, downward sausage wave energy flux, $E_{\|}^{(d)}$, plotted as $z\rightarrow\infty$ (dashed) for reference, thus does not relate to the horizontal scale and is a constant for each mode order and frequency.  Each mode order is plotted with a different colour, see figure \ref{fig:disp} for the colour scheme.}\label{fig:EuEd}
\end{figure}

The aim is to understand how MHD tube waves can transport energy from $p$ modes in the convection zone down into the deep interior and up into the upper photosphere/lower chromosphere.  In Appendix \ref{EF:sausage} we derive expressions for the upward, $E^{(u)}_{\|}$, and downward, $E^{(d)}_{\|}$, energy flux of sausage waves driven by $f$ and $p$ modes in a two-region, polytropic and isothermal, atmosphere.  We quantify the energy loss of $f$ and $p$ modes as a \textit{damping rate}, $\Gamma$, and \textit{absorption coefficient}, $\alpha$, which are explicitly calculated in Appendix \ref{absdamp} from the wave energy flux of sausage waves and the energy contained in $f$ and $p$ modes for the given model.\\
\\
In figures \ref{fig:damp4H} and \ref{fig:alpha4H} we present damping rates and absorption coefficients for three values of the plasma-$\beta=0.1$, $1$, $10$.  The $\beta$ dependence of both the damping and of absorption vary non-linearly, with a peak at around $\beta=1$ and decreasing in magnitude with increasing and decreasing $\beta$ from this peak.  The upward and downward wave energy flux are calculated as $z\rightarrow\infty$ and $z\rightarrow-\infty$ respectively.  No wave energy escapes the upper domain into the overlying atmosphere since the sausage waves are evanescent in the frequency regime we are restricted to (see figure \ref{fig:disp}  below dot-dashed line).  Therefore all wave energy which escapes the acoustic cavity in the model is through the lower domain.  This property is similar to that of a model with a stress-free condition imposed at the photosphere, except that additional driving can occur above the photosphere and the phase shift of the reflected wave may differ.\\
\\
The low frequency portion of the damping rates and absorption coefficients are in very good agreement to previous studies where a stress-free boundary condition was applied at the model photosphere ($z=-z_{0}$) \citep[see][]{2008HindmanJ,2009JainHBB}.  In this frequency regime the upper turning point of both sausage waves and $p$ modes are well below the interface, thus, the isothermal region behaves very similar to a stress-free boundary at $z=-z_{0}$ (see figure \ref{fig:dampCz0392}).  This also implies that for low frequencies most of the $p$-mode energy lost to the tube escapes down the tube and very little escapes into the isothermal atmosphere, which is apparent when the energy fluxes $E_{\|}^{(u)}$ and $E_{\|}^{(d)}$ are investigated in detail (see \S\ref{EF:sausage}).  In figure \ref{fig:EuEd} we plot the upward sausage wave energy flux, $E_{\|}^{(u)}$, (equation (\ref{EF:u:iso})) as a function of height and the corresponding value of $E_{\|}^{(d)}$ as a reference for how large the upward energy flux is.  The solid curve in figure \ref{fig:EuEd} shows that the upward wave energy flux decays with height, but, the rate of decay decreases with frequency.  For frequencies above $3$ mHz a significant amount of energy flux remains in the tube, even a few scale heights above the interface, this energy flux eventually decays to zero and does not leave the domain i.e., does not contribute to the damping and absorption of $p$ modes.  The sausage waves are evanescent in the isothermal region thus their wave energy flux is not being carried by free oscillations.  The flux is the result of driven oscillations from below and the flux decreases with height because the $p$-mode driver (also evanescent) acts as a sink in the isothermal region removing energy form the tube.  Although caution must be taken for frequencies too close to the cut-off frequency, since very close to this frequency the $p$-mode eigenfunction and, hence the wave driver, has a skin depth that is insufficiently short to ensure that the excitation remains finite with height. This manifests as divergent behaviour in $E_{\|}^{(u)}$ in the limit $z\to\infty$. Explicitly, the term $\mathcal{S}$ in equation (\ref{EF:u:iso}) diverges with height because of its dependence on $\sim\exp(\frac{1}{2}[i\lambda_{z,e}+i\lambda_{\|}+(a+1)/2])$.  This divergent behaviour is clearly unphysical and can be avoided if we ensure that $|k_{z}'|+|k_{\|}|>\frac{1}{4H}$, or for the parameters used here, if we restrict our attention to frequencies below $4.23$ mHz.\\
\\
The inclusion of an upper isothermal atmosphere allows us to take into account the coupling of all waves between the two regions which is an arduous procedure but required in order to model the reflective properties of the upper atmosphere with even minimal fidelity.  In figure \ref{fig:dampCz0392} we compare damping rates for our two-region (infinite domain) model with truncated polytrope models with stress-free and maximal-flux boundary conditions applied at the interface ($z=-z_{0}$) \citep{1996BogdanHCC,2008HindmanJ,2009JainHBB}.  As noted earlier the low frequency behaviour of the two-region model is similar to that of the stress-free case; the isothermal atmosphere reflects all waves back down the tube in a similar manner as in the stress-free case.  For higher frequencies the damping of $p$ modes decrease compared to the stress-free case, this is due to the mechanism discussed earlier, that the $p$ modes act as a sink in the isothermal atmosphere extracting energy out of sausage waves.  In the stress-free model the upward propagating wave is forced to reflect at the interface and thus propagates down the tube and this energy loss is measured at the lower boundary as a combination of downward and upward reflected propagating waves.  In the present two-region case, this upward propagating wave in the polytrope is no longer reflecting in the same manner at the interface and instead becomes evanescent very close to the interface and a portion of its wave energy is transported into the isothermal region.  This energy flux then decreases to zero with height and thus we measure a zero contribution to the escaping wave energy flux, reducing the damping rate at higher frequencies where this effect becomes more and more dominant.  The maximal-flux condition (introduced in \citet{2008HindmanJ}) is vital in providing an upper limit on the amount of energy lost from $f$ and $p$ modes to magnetic tube waves.  However it does not provide insight into the nature of the interaction of tube waves with a realistic overlying atmosphere (figure \ref{fig:dampCz0392} dashed lines).

\subsection{Conclusion}
In this paper we have presented a calculation for damping rates and absorption coefficients of solar $p$ modes due to the excitation of sausage tube waves in a two-region (isothermal-polytropic) atmosphere.  By including an isothermal region above the $p$-mode cavity we have shown that the linewidths (or damping rates) of solar $f$ and $p$ modes are sensitive to a change in the reflective properties of the surface.\\
\\
Our two-region atmosphere model demonstrates the importance of coupling between the interior to the upper atmosphere for the near surface behaviour of the solar acoustic oscillations and the magnetic sausage tube waves.  The comparison of damping rates and absorption coefficients with previous such studies demonstrates that a stress-free boundary condition applied at the photosphere gives a good approximation, at low frequencies, but should be considered with caution as one approaches closer to the acoustic cut-off frequency.\\
\\
For the absorption coefficients, it has also been previously noted that at low frequencies below $3$ mHz the theoretical estimates are qualitatively similar to the observations \citep{2009JainHBB} which is consistent with the results presented here.  Above $3$$-$$4$ mHz the observations underestimate the absorption of $p$-mode energy and show a rapid fall-off of the absorption coefficient which is likely to be due to the existence of high-frequency halos surrounding active regions \citep{2002JainH} and due to acoustic jacket modes in the $f$- and $p$-mode wavefields \citep{1995BogdanC,2013Cally}.

This work is supported by STFC (UK).  BWH also acknowledges NASA grants NNX09AB04G, NNX14AC05G and NNZ14AG05.

\appendix

\section{Wavefields in the non-magnetic surroundings}\label{A:wavefield:nm}
In this section we shall derive the wavefields for both the polytropic and isothermal atmospheres.  At the interface $z=-z_{0}$ the vertical component of the displacement perturbation $\xi_{z,e}$, and the Lagrangian pressure perturbation $\delta p_{e}=-\gamma p_{e}\grad\cdot\vec{\xi_{e}}$ are both (assumed to be) continuous.

\subsection{Polytropic wavefield}
In the field-free polytropic atmosphere we follow the procedure adopted by \citet{1996BogdanHCC,2008HindmanJ,2011GascoyneJH}.  Thus, the wave perturbations $p_{e}'$, $\rho_{e}'$, $\vec{\xi_{e}}$, in pressure, density and fluid displacement respectively, can be written in terms of a displacement potential $\Phi$,
\begin{equation}\label{Phi:form}
\vec{\xi_{e}}=\grad\Phi,\hspace{3mm}p_{e}'=-\rho_{e}\frac{\partial^2\Phi}{\partial t^2},\hspace{3mm}\rho_{e}'=-\frac{\rho_{e}}{c^2}\frac{\partial^2\Phi}{\partial t^2},
\end{equation}
which together form the wave equation \citep[see][]{1932Lamb},
\begin{equation}\label{wave:eqn:poly}
    \frac{\partial^2\Phi}{\partial t^2}=c^2\grad^2\Phi-g\frac{\partial\Phi}{\partial z}.
\end{equation}
The incident acoustic wavefield expressed by equation (\ref{wave:eqn:poly}) supports plane wave solutions of the form,
\begin{equation}\label{wave:sol:poly}
    \Phi(\vec{x},t)=z_{0}\mathcal{A}e^{i(k_{x}x-\omega t)}Q(z),
\end{equation}
where, $\mathcal{A}$, is the complex wave amplitude, $\omega$, the temporal frequency, $k_{x}$, the horizontal wavenumber and $Q(z)$, the vertical eigenfunction.  Thus, (\ref{wave:eqn:poly}) yields,
\begin{equation}\label{Q(w)}
    Q(w)=w^{-(\mu+1/2)}W_{\kappa,\hspace{0.5mm}\mu}(w),
\end{equation}
where $W_{\kappa,\hspace{0.5mm}\mu}$ is the Whittaker function of the second kind.  Also,
\begin{equation}\label{nondim:par}
w=-2k_{x}z,\quad\nu^2=\frac{a\omega^2z_{0}}{g},\quad\kappa=\frac{\nu^2}{2k_{x}z_{0}},\quad\mu=\frac{(a-1)}{2},\quad\lambda=2k_{x}z_{0}.
\end{equation}

\subsection{Isothermal wavefield}\label{Iso:wavefield}
In the field-free isothermal atmosphere the linearised hydrostatic equations can be manipulated to form the governing wave equation for the vertical displacement perturbation as,
\begin{equation}\label{iso:gov}
    \frac{\partial^2\vec{\xi_{e}}}{\partial t^2}=c^2\grad\left(\grad\cdot\vec{\xi_{e}}\right)+(\gamma-1)\vec{g}\left(\grad\cdot\vec{\xi_{e}}\right)+\grad\left(\vec{g}\cdot\vec{\xi_{e}}\right).
\end{equation}
Since this partial differential equation has constant coefficients we Fourier analyse in time and space i.e., $\vec{\xi_{e}}=\vec{A}\exp\left[i\left(\vec{k}\cdot\vec{x}-\omega t\right)\right]$.  If we refer to figure \ref{fig:atmos} the isothermal atmosphere is situated on top of the polytropic atmosphere with its base at $z=-z_{0}$, thus substituting $z=z+z_{0}$ and denoting $k_{z}=-i\frac{1}{2H}+k_{z}'$ (see equation (\ref{kz'})) the displacement $\vec{\xi_{e}}$ can be written as,
\begin{equation}
\vec{\xi_{e}}=\vec{A}e^{(z+z_{0})/(2H)}e^{i\left(k_{z}'(z+z_{0})+k_{x}x-\omega t\right)}.
\end{equation}

\subsection{Matching conditions across the interface}
The continuity of Lagrangian pressure perturbation at the interface yields the following relation,
\begin{equation}\label{freq:eigen}
\left[\nu^2a-2\kappa^2(i\lambda_{z,e}+a+1)\right]W_{\kappa,\hspace{0.5mm}\mu}(\lambda)+\kappa(2\kappa-a)(i\lambda_{z,e}-a+1)W_{\kappa,\hspace{0.5mm}\mu+1}(\lambda)=0,
\end{equation}
where,
\begin{equation}\label{lamz}
\lambda_{z,e}=2k_{z}'z_{0}=\sqrt{\frac{\lambda}{\kappa}\left(4\kappa^2+a-\nu^2\right)-(a+1)^2}.
\end{equation}
Note that in order to solve equation (\ref{freq:eigen}) we require that $\lambda_{z,e}$ is purely imaginary, thus we only consider trapped $p$-mode solutions.  We couple the propagating waves in the polytrope on to evanescent waves in the isothermal atmosphere, thus we only consider the frequencies below the acoustic cut-off frequency which is given by $\lambda_{z,e}^2<0$.\\
\\
After solving equation (\ref{freq:eigen}) for the eigenfrequencies, using the non-dimensional parameters (\ref{nondim:par}) the vertical component of the eigenfunctions can be calculated as,
\begin{equation}\label{xi:iso}
\xi^{\text{\tiny (I)}}_{z,e}=-z_{0}\frac{\mathcal{A}}{z_{0}}\lambda\left.\frac{dQ}{dw}\right|_{w=\lambda}e^{\frac{1}{2}(i\lambda_{z,e}+a+1)\left(1-\frac{w}{\lambda}\right)},\quad w\leq\lambda,
\end{equation}
\begin{equation}\label{xi:poly}
\xi^{\text{\tiny (P)}}_{z,e}=-z_{0}\frac{\mathcal{A}}{z_{0}}\lambda \frac{dQ}{dw},\quad w\geq\lambda,
\end{equation}
where we have used the notation $\xi^{(I)}_{z,e}$ and $\xi^{(P)}_{z,e}$ for the isothermal and polytropic vertical displacements, respectively.  Above we have omitted the Fourier components $e^{i(kx-\omega t)}$, which we shall continue to do so for brevity but still take into account the $t$ and $x$ derivatives when encountered.

\subsection{Energy contained in $f$- and $p$-mode oscillations}\label{sec:En}
The wave energy density of $f$ and $p$ modes for the prescribed model is given by \citep[see][]{1974BrayL},
\begin{equation}\label{energy:den}
\mathcal{W}=\frac{1}{4}\rho_{e}\left|\frac{\partial\vec{\xi}_{e}}{\partial t}\right|^2+\frac{|p'_{e}|^2}{4\rho_{e}c^2}-\frac{1}{4}\frac{g^2}{c^2}\rho_{e}\left(1+\frac{c^2}{g\rho_{e}}\frac{d\rho_{e}}{dz}\right)|\xi_{z,e}|^2.
\end{equation}
For a neutrally stable atmosphere the last term, which corresponds to the gravitational energy, vanishes.  Taking the time average, integrating over depth and multiplying by $4\pi R^2_{\odot}$ gives the energy contained in $f$- and $p$-mode oscillations which we shall denote $E_{n}$.  The polytropic segment of this integral is given by \citet{2008HindmanJ},
\begin{equation}\label{EPn}
E^{(P)}_{n}=4\pi R^2_{\odot}\frac{g\rho_{0}z_{0}^2}{4a}\lambda^2\frac{|\mathcal{A}|^2}{z_{0}^2}\mathcal{N}_{n}^{(1)},
\end{equation}
where,
\begin{equation}\label{Nn1}
\mathcal{N}_{n}^{(1)}=\frac{\kappa^2}{\lambda^a}\int^{\infty}_{\lambda}dw\frac{w^a}{\kappa}\left[\left(\frac{dQ(w)}{dw}\right)^2+\left(\frac{1}{4}+\frac{\kappa}{w}\right)Q^2(w)\right].
\end{equation}
The isothermal contribution to the wave energy density is calculated as \citep[see also,][]{1997Hasan},
\begin{equation}\label{hasan:W}
\mathcal{W}^{(I)}=\frac{p_{e}}{4}\left\{\frac{\gamma}{c^2}(\omega^2+\omega^2_{\text{\tiny BV}})|\xi_{z,e}|^2+\frac{1}{\gamma}\left(1+\frac{c^2k_{x}^2}{\omega^2}\right)\left|\frac{p'_{e}}{p_{e}}\right|^2\right\}.
\end{equation}
Integrating equation (\ref{hasan:W}) over the domain (which is from the interface $z=-z_{0}$ to $z=\infty$ or $w=\lambda$ to $w=-\infty$) and multiplying by our model surface area we obtain,
\begin{equation}\label{EIn}
E^{(I)}_{n}=4\pi R^2_{\odot}\frac{g\rho_{0}z^2_{0}}{4a}\lambda^2\frac{|\mathcal{A}|^2}{z^2_{0}}\mathcal{N}_{n}^{(2)},
\end{equation}
where,
\begin{equation}\label{Nn2}
\mathcal{N}_{n}^{(2)}=\left\{\nu^2+a+\left(1+\frac{\nu^2}{4\kappa^2}\right)\left[\frac{2\kappa^2(i\lambda_{z,e}-a+1)}{\nu^2-4\kappa^2}\right]^2\right\}\left(\frac{1}{-i\lambda_{z,e}}\right)\left(\left.\frac{dQ}{dw}\right|_{w=\lambda}\right)^2.
\end{equation}
The total energy contained in the $f$- and $p$-mode oscillations for the two-region model is $E_{n}=E^{(P)}_{n}+E^{(I)}_{n}$.

\subsection{Lateral energy flux of $f$- and $p$-mode oscillations}
The external (to the flux tube) wavefield in both the polytropic and isothermal regions can be converted from Cartesian ($x$,$y$,$z$) to cylindrical polar coordinates ($r$,$\phi$,$z$) with origin centred on the flux tube (i.e., $r=0$) using the Jacobi-Anger expansion.  Thus, in both regions the wavefields functional dependence on $r$, $\phi$ and $t$ is,
\begin{equation}\label{jacobi:anger}
f(r,\phi,t)=\frac{1}{2}e^{i(m\phi-\omega t)}H^{(2)}_{m}(kr).
\end{equation}
The energy flux of $f$ and $p$ modes for the prescribed model is given by \citep[see][]{1974BrayL},
\begin{equation}
\vec{F}=p'_{e}\vec{v}_{e}.
\end{equation}
Taking the time average and integrating over a cylinder of radius $r$ gives the lateral energy flux of $f$- and $p$-mode oscillations which we shall denote $F_{n}$.  The polytropic component of the lateral energy flux i.e., from $z=-\infty$ to $z=-z_{0}$,  can be written as \citep[see][]{2009JainHBB},
\begin{equation}\label{FPn}
F^{(P)}_{n}=-\frac{|\mathcal{A}|^2}{z_{0}^2}\rho_{0}z_{0}^2\left(\frac{gz_{0}}{a}\right)^{3/2}\mathcal{H}^{(1)}_{n},
\end{equation}
where,
\begin{equation}\label{Hn1}
\mathcal{H}^{(1)}_{n}=\frac{\nu^3}{2}\left(\frac{\kappa}{\nu^2}\right)^{a+1}\int^\infty_{\frac{\nu^2}{\kappa}}dw~w^aQ^2(w).
\end{equation}
We now consider the calculation of the isothermal component of the lateral energy flux i.e., from $z=-z_{0}$ to $z=\infty$,
\begin{equation}\label{iso:Fr}
<F^{(I)}_{r}>=\int^{2\pi}_{0}rd\phi\int^{\infty}_{-z_{0}}\frac{1}{4}(p'_{e}v'^*_{r,e}+p'^*_{e}v'_{r,e})dz.
\end{equation}
Firstly we calculate $v'_{r,e}=-i\omega\xi_{r,e}$ as,
\begin{equation}\label{v'_re}
v'_{r,e}=-\frac{2i}{\omega}\frac{\partial}{\partial r}\left(\frac{p'_{e}}{\rho_{e}}\right).
\end{equation}
In cylindrical polar coordinates $p'_{e}$ can be written using (\ref{jacobi:anger}) in the form,
\begin{equation*}
p_{e}'=-\gamma p_{e}\lambda \left.\frac{dQ}{dw}\right|_{w=\lambda}\frac{\mathcal{A}}{2z_{0}}\left[\frac{2\kappa^2(i\lambda_{z,e}-a+1)}{\nu^2-4\kappa^2}\right]
\end{equation*}
\begin{equation}\label{p'_e}
\times e^{\frac{1}{2}\left(i\lambda_{z,e}+a+1\right)\left(1-\frac{w}{\lambda}\right)}H^{(2)}_{m}(kr)e^{i(m\phi-\omega t)}.
\end{equation}
Substituting equations (\ref{v'_re}) and (\ref{p'_e}) into (\ref{iso:Fr}) and using the Wronskian for the Hankel functions we obtain,
\begin{equation}\label{FIn}
F^{(I)}_{n}=-\frac{|\mathcal{A}|^2}{z_{0}^2}\rho_{0}z_{0}^2\left(\frac{gz_{0}}{a}\right)^{3/2}\mathcal{H}^{(2)}_{n},
\end{equation}
where,
\begin{equation}\label{Hn2}
\mathcal{H}^{(2)}_{n}=\nu^3\left(\frac{1}{-i\lambda_{z,e}}\right)\left(\left.\frac{dQ}{dw}\right|_{w=\lambda}\right)^2\left[\frac{2\kappa(i\lambda_{z,e}-a+1)}{\nu^2-4\kappa^2}\right]^2.
\end{equation}
The time averaged lateral energy flux of solar $f$- and $p$-mode oscillations for the two-region model is $F_{n}=F^{(P)}_{n}+F^{(I)}_{n}$.

\section{Governing equations for the sausage tube waves driven by p-mode buffeting in a two-region atmosphere}\label{GE:sausage}
Under the thin flux tube approximation we derive a set of MHD equations for an inviscid and infinitely conducting medium.  On linearising these equations and assuming a time dependence consistent with the external perturbations we can describe the interaction of the field-free medium with the thin flux tube for the sausage mode in both the polytropic and isothermal atmosphere.  Thus using the formulation of \citet{2008HindmanJ,2009JainHBB,2011GascoyneJH} for the polytropic atmosphere and \citet{1997Hasan} for the isothermal atmosphere we obtain the governing equations for the propagation of sausage waves for our prescribed two-region atmosphere,
\begin{equation}
\left[c^2_{\text{\tiny T}}\frac{\partial^2}{\partial z^2}-\frac{\gamma g}{2}\frac{c^2_{\text{\tiny T}}}{c^2}\frac{\partial}{\partial z}+\omega^2+N_{g}\right]\xi_{z}=\frac{c^2_{\text{\tiny T}}}{\rho V^2_{\text{\tiny A}}}\left(\frac{\partial}{\partial z}+\frac{g}{c^2}\right)p_{e}',
\end{equation}
where $V^2_{\text{\tiny A}}=B^2/(4\pi\rho)$ and $c^2_{\text{\tiny T}}=c^2V^2_{\text{\tiny A}}/(1+c^2/V^2_{\text{\tiny A}})$ are the squares of Alfv\'{e}n speed and tube speed respectively.  The governing equations for the propagation of sausage waves in the two regions are essentially the same apart from the term $N_{g}$ ($=-\omega_{\text{\tiny BV}}^2\left(c^2+\frac{\gamma}{2}V^2_{\text{\tiny A}}\right)/(c^2+V^2_{\text{\tiny A}})$) which permits the propagation of gravity waves in the isothermal layer; this quantity is zero for the polytrope which is in adiabatic equilibrium and therefore, gravity waves cannot propagate.  For the two atmospheres we obtain the following solutions to equation (\ref{gov:eqn}),
\begin{equation}\label{xi:I}
\xi^{(I)}_{\|}=-\mathcal{A}\left[B_{1}e^{b_{1}\left(1-s\right)}+B_{2}e^{b_{2}\left(1-s\right)}+\mathcal{K}e^{b_{3}\left(1-s\right)}\right]
\end{equation}
\begin{equation}\label{xi:P}
\xi^{(P)}_{\|}=-\frac{i\pi}{2}\mathcal{A}\left\{\psi_{\|}(s)\left[\Omega+\mathcal{J}^*(s)\right]+\psi_{\|}^*(s)\left[\mathcal{I}-\mathcal{J}(s)\right]\right\}
\end{equation}
where $*$ denotes the complex conjugate and,
\begin{equation*}
b_{1}=\frac{1}{4}(2i\lambda_{\|}+a+1),\quad b_{2}=-\frac{1}{4}(2i\lambda_{\|}-a-1),\quad b_{3}=\frac{1}{2}\left(i\lambda_{z,e}+a+1\right),
\end{equation*}
\begin{equation}
\lambda_{\|}=2k_{\|}z_{0}=\sqrt{2(\nu^2-a)(2+\gamma\beta)-\frac{(a-3)^2}{4}}
\end{equation}
\begin{equation}
\mathcal{K}=\lambda\left.\frac{dQ}{dw}\right|_{w=\lambda}\frac{2\gamma(1+\beta)(\nu^2-a)}{\left\{\left(i\lambda_{z,e}+\frac{a+1}{2}\right)^2+\lambda^2_{z}\right\}},
\end{equation}
\begin{eqnarray}\label{psi}
 \psi_{\|}(s)&=&s^{-\mu/2}H_{\mu}^{(1)}(2\nu\sqrt{\epsilon s}),\\\label{Integral:1:ch2}
\mathcal{J}(s)&=&\int^{s}_{1}r^{\mu+1}\psi_{\|}(r)f(r)dr,\\\label{Integral:3:ch2}
\mathcal{I}&=&\int^{\infty}_{1}r^{\mu+1}\psi_{\|}(r)f(r)dr,\\
f(s)&=&-\frac{(1+a)(\beta+1)}{2a}\frac{\nu^2}{s}\frac{d Q(s)}{d s}.
\end{eqnarray}
We have also introduced the dimensionless parameters $s=-z/z_{0}$ and $\epsilon=(2a+\beta(1+a))/(2a)$.  Assuming a finite energy density at the upper boundary ($s\rightarrow\infty$) we set $B_{2}=0$.

\subsection{Matching conditions for the sausage tube waves at the interface}\label{match:sausage}
At the interface $s=1$ we require the two solutions $\xi^{(I)}_{\|}$ and $\xi^{(P)}_{\|}$ (equations (\ref{xi:I}) and (\ref{xi:P}) respectively) to be equal, thus we obtain a relation for $B_{1}$,
\begin{equation}
B_{1}=\frac{i\pi}{2}\left(H_{\mu}^{(1)}(2\nu\sqrt{\epsilon})\Omega+H_{\mu}^{(2)}(2\nu\sqrt{\epsilon})\mathcal{I}\right)-\mathcal{K}.
\end{equation}
The only other parameter needing to be specified is the constant $\Omega$, which is calculated by ensuring that the Lagrangian pressure perturbation inside the tube is continuous across the interface.  We obtain the following result,
\begin{equation}
\Omega=\frac{i}{\pi\mathcal{R}_{1}}\bigg[\left(\gamma\nu^2Q_{0}-\Lambda\right)(\beta+1)+2(b_{1}-b_{3})\mathcal{K}\bigg]-\frac{\mathcal{R}_{2}}{\mathcal{R}_{1}}\mathcal{I}
\end{equation}
where,
\begin{equation}
\mathcal{R}_{1}=\nu\sqrt{\epsilon}H_{\mu+1}^{(1)}(2\nu\sqrt{\epsilon})-b_{1}H_{\mu}^{(1)}(2\nu\sqrt{\epsilon}),
\end{equation}
\begin{equation}
\mathcal{R}_{2}=\nu\sqrt{\epsilon}H_{\mu+1}^{(2)}(2\nu\sqrt{\epsilon})-b_{1}H_{\mu}^{(2)}(2\nu\sqrt{\epsilon}).
\end{equation}
In figure \ref{fig:eigenf3n2b1} we plot the displacement for the incident $p_{2}$-mode with its corresponding excited sausage wave.  Inset plot for $s>1$ shows the familiar polytropic solutions for the $p_{2}$-mode and sausage wave.  The $s<1$ isothermal region shows the evanescent tail which grows with height due to the decrease in density.  Figure \ref{fig:eigenf3n2b1} clearly shows the smooth matching of the wave solutions across the interface $s=1$.

\section{Energy flux of sausage tube waves}\label{EF:sausage}
The wave energy flux at any point along the tube is given by the following expression \citep{1974BrayL},
\begin{equation}\label{energy:flux}
\vec{F}=p'\vec{v}+\frac{1}{4\pi}\left[\left(\vec{B}\cdot\vec{B}'\right)\vec{v}-\left(\vec{B}'\cdot\vec{v}\right)\vec{B}\right],
\end{equation}
where $p'$, $\vec{v}$ and $\vec{B}'$ are the pressure perturbation, velocity vector and magnetic field perturbation vector respectively.  Reducing equation (\ref{energy:flux}) to obtain the parallel component (with respect to the tube axis),
\begin{equation}\label{pre:F}
F_{\|}=p_{e}'\frac{\partial\xi_{\|}}{\partial t}-\frac{BB_{\|}'}{4\pi}\frac{\partial\xi_{\|}}{\partial t}.
\end{equation}
The perturbed magnetic field $B_{\|}'$ can be obtained from the derivation of the governing equation (\ref{gov:eqn}) and takes the same form in both layers,
\begin{equation}\label{B'}
\frac{BB_{\|}'}{4\pi}=\frac{2}{(2+\gamma\beta)}p'_{e}+\frac{\gamma\beta}{(2+\gamma\beta)}\frac{B^2}{4\pi}\frac{\partial\xi_{\|}}{\partial z}-g\left(\rho-\frac{\gamma\beta}{(2+\gamma\beta)}\rho_{e}\right)\xi_{\|}.
\end{equation}
Substituting equations (\ref{xi:iso}), (\ref{xi:poly}) and (\ref{B'}) into equation (\ref{pre:F}) and taking the time average, after substantial manipulation, we obtain the energy flux of sausage tube waves escaping from the magnetic flux tube as,
\begin{equation}\label{EF:u:iso}
E_{\|}^{(u)}=\frac{\gamma\beta}{(2+\gamma\beta)}\frac{2g\rho_{0}z^2_{0}\omega A_{0}}{4(a+1)(\beta+1)}\frac{|\mathcal{A}|^2}{z_{0}^2}\bigg[\text{Re}(\lambda_{\|})|B_{1}|^2+\mathcal{S}\bigg],
\end{equation}
\begin{equation}\label{EF:d:poly}
E_{\|}^{(d)}=-\frac{\gamma\beta}{(2+\gamma\beta)}\frac{\pi g\rho_{0}z_{0}^2\omega A_{0}}{4(a+1)(\beta+1)}\frac{|\mathcal{A}|^2}{z_{0}^2}|\Omega+\mathcal{I}^*|^2,
\end{equation}
where,
\begin{equation}
\mathcal{S}=e^{\frac{1}{2}i\lambda_{z,e}(1-s)}\left.\bigg[2\mathcal{K}\text{Im}\left(b_{1}B_{1}e^{b_{1}(1-s)}\right)-\{\Lambda(\beta+1)+2b_{3}\mathcal{K}\}\text{Im}\left(B_{1}e^{b_{1}(1-s)}\right)\bigg]\right|_{s\rightarrow-\infty},
\end{equation}
\begin{equation}\label{Lambda}
\Lambda=-\gamma\lambda Q'_{0}\left[\frac{2\kappa^2(i\lambda_{z,e}-a+1)}{\nu^2-4\kappa^2}\right].
\end{equation}
Note above that `Re' and `Im' denote the real and imaginary parts respectively of the corresponding complex quantities inside the parenthesis.  $E_{\|}^{(u)}$ and $E_{\|}^{(d)}$ denote the wave energy flux calculated at $z\rightarrow\infty$ ($s\rightarrow-\infty$) and $z\rightarrow-\infty$ ($s\rightarrow\infty$) respectively.  We have also multiplied the energy fluxes by the cross-sectional area of the tube,
\begin{equation}
A(s)=\left\{
  \begin{array}{cc}
    A_{0}e^{(\mu+1)(1-s)} & \quad \text{$s\leq 1$ isothermal}\\
    A_{0}s^{-(\mu+1)} & \quad \text{$s\geq 1$ polytrope}.\\
  \end{array} \right.
\end{equation}
We find that $E_{\|}^{(u)}$ is zero for the frequency range under consideration.

\subsection{$p$-Mode damping rates and absorption coefficients}\label{absdamp}

We define the damping rate $\Gamma$ for a single tube as \citep[see also][]{2008HindmanJ},
\begin{equation}\label{pre:gamma}
\Gamma=-\frac{1}{2\pi}\frac{E_{\|}}{E_{n}}=-\frac{1}{2\pi}\frac{E^{(u)}_{\|}+E^{(d)}_{\|}}{E^{(P)}_{n}+E^{(I)}_{n}}.
\end{equation}
Substituting equations (\ref{EPn}), (\ref{EIn}), (\ref{EF:u:iso}) and (\ref{EF:d:poly}) into (\ref{pre:gamma}) we obtain,
\begin{equation}\label{Gamma}
\Gamma=\frac{\beta}{4(\beta+1)\epsilon}\frac{\omega A_{0}}{4\pi R^2_{\odot}\lambda^2}\frac{\left|\Omega+\mathcal{I}^*\right|^2}{\mathcal{N}_{n}},
\end{equation}
where $\mathcal{N}_{n}=\mathcal{N}^{(1)}_{n}+\mathcal{N}^{(2)}_{n}$ (see equations (\ref{Nn1}) and (\ref{Nn2})).  Multiplying by the number of tubes on the solar surface $N=4\pi R^2_{\odot}f/A_{0}$ where $f$ is the magnetic filling factor (fraction of the solar surface the tubes occupy) we compute the total damping rate for $f$ and $p$ modes.\\
\\
We follow the same methodology as in \citet{2009JainHBB,2011GascoyneJH} and define the absorption coefficient $\alpha$ for a single tube as,
\begin{equation}\label{pre:alpha}
\alpha=\frac{E_{\|}}{F_{n}}=\frac{E^{(u)}_{\|}+E^{(d)}_{\|}}{F^{(P)}_{n}+F^{(I)}_{n}},
\end{equation}
Substituting equations (\ref{FPn}), (\ref{FIn}), (\ref{EF:u:iso}) and (\ref{EF:d:poly}) into (\ref{pre:alpha}) after simplification we obtain,
\begin{equation}\label{alpha}
\alpha=\frac{\pi\beta}{4(\beta+1)\epsilon}\frac{A_{0}}{z_{0}^2}\frac{\left|\Omega+\mathcal{I}^*\right|^2}{\mathcal{H}_{n}},
\end{equation}
where $\mathcal{H}_{n}=\mathcal{H}^{(1)}_{n}+\mathcal{H}^{(2)}_{n}$ (see equations (\ref{Hn1}) and (\ref{Hn2})).


\end{document}